\documentclass[twocolumn,showpacs,preprintnumbers,amsmath,amssymb]{revtex4}
\usepackage{graphicx}% Include figure files
\usepackage{dcolumn}% Align table columns on decimal point
\usepackage{bm}% bold math
\begin{document}

\title{Supersymmetric quantum mechanics to study the 16.8 MeV resonance state of $^{9}$B}
\author{S. K. Dutta$^1$}
\author{D. Gupta$^2$}
\email{dhruba@jcbose.ac.in}
\author{Swapan K. Saha$^2$}
\altaffiliation[Present Address:] {121/2 Manicktala Main Road, Kolkata 700054}
%{\footer{dhruba@jcbose.ac.in}}
\affiliation{$^1$Department of Physics, B.G. College, Berhampore, Murshidabad 742101, India}
\affiliation{$^2$Department of Physics, Bose Institute, 93/1 A.P.C. Road, Kolkata 700009, India}
\date{\today}
\begin{abstract}
The study of a high-lying resonant state of $^{9}$B is carried out using 
supersymmetric quantum mechanics (SQM). The resulting isospectral potentials are very deep 
and narrow and the generated wave functions identify the resonance at 16.84 MeV with a width of 
69 keV. The present work shows that SQM can be successfully applied for detection of 
high-lying resonances in unstable nuclei. 
\end{abstract}
\maketitle

\section{Introduction}
\noindent 
With the development of new and upgraded rare isotope beam facilities worldwide, one can now 
carry out sophisticated experiments involving resonances in unstable nuclei~\cite{SN13,RI19}.
This opens up immense opportunities to pursue pressing problems in nuclear astrophysics. 
The cosmological lithium problem~\cite{CY08,BO10} is one such decades old and yet 
unresolved problem, where there is a pronounced abundance anomaly for $^7$Li between the 
observation and prediction of the Big Bang Nucleosynthesis theory. It has been argued that
resonance enhancement through a high-lying state in the $^9$B nucleus might give a
clue to the solution to this long standing problem~\cite{RI19,CY08,BO10,AN05,CH11}. We have
developed a robust theoretical framework using supersymmetric quantum mechanics (SQM) to generate 
the resonant states and their wave functions for unstable and unbound nuclei with 
excellent results~\cite{DU14,DU18}. In the present work we study the high-lying resonance
of $^9$B in this framework. 

\section{Theory}
\noindent 
Earlier, using SQM we have studied low-lying resonances in unstable and unbound nuclei~\cite{DU14,DU18}. However, we anticipated that detection of high-lying resonances in these nuclei would
require very deep isospectral potentials resulting in serious challenges as described below.\\

\noindent 
In our present work, using SQM, a high-lying resonant state of $^{9}$B is investigated. 
In essence, this tests the effectiveness of our theoretical procedure for high-lying resonant 
states of unstable nuclei. The challenge in detecting such resonant states results from the
shallowness of the two-body potential well (d + $^{7}$Be for $^{9}$B), followed by a very low 
and wide barrier. Thus, for a finite barrier height, a system may temporarily be trapped inside 
the shallow well, when its energy is close to the resonance energy. In principle one can find 
quasi-bound states in such a shallow potential. However there is a large probability to tunnel 
out through the barrier which gives rise to broad resonance width. For high-lying resonant
states, the probability of tunnelling becomes so high that it makes the accurate detection 
of such states practically impossible. In order to circumvent this problem, we resort to the 
very successful procedure adopted by us earlier in the detection of low-lying resonant states.\\

\noindent
Since $^9$B is formed by the fusion of $^7$Be with a deuteron in the context of the lithium 
problem~\cite{KI11},
we study $^{9}$B in a two-body model consisting of a $^{7}$Be core and a deuteron.
The two-body potential $v(r)$ is generated microscopically in a double folding model 
using densities of the deuteron and $^{7}$Be along with the density dependent M3Y (DDM3Y) 
effective interaction. The details of the framework are explained in~\cite{DU14,BA03}. 
The densities of deuteron and $^{7}$Be used in the present work are obtained from variational 
Monte Carlo calculations using the Argonne v18 two-nucleon and Urbana X three-nucleon 
potentials (AV18+UX)~\cite{WI91}. Earlier, the DDM3Y effective interaction was succesfully 
used to describe nuclear matter~\cite{BA04}, radioactivity~\cite{BA05}, scattering~\cite{GU06} 
as well as resonances in unstable~\cite{DU14} and unbound nuclei~\cite{DU18}. In the present 
work, it is used to study high-lying resonant states in unstable nuclei in the context of 
astrophysical problems.\\

\noindent 
From the above microscopic potential $ v(r)$, inclusive of the centrifugal barrier, 
we use SQM to generate a family of isospectral 
potentials (IP) involving an arbitrary parameter $\lambda$. These potentials may appear quite 
different but they have exactly the same energy spectrum as the original one. The idea of 
isospectral potential has been extended by Pappademos {\it et al}~\cite{PA93} to scattering 
states with positive energy in the continuum. While the wave functions in the continuum are 
non-normalizable, following Pappademos et al., one can construct a normalizable wave function
at a selected energy. This represents a bound state in the continuum (BIC). The BIC is a 
solution of the Schr\"odinger equation with an isospectral potential $\hat{v}(r;\lambda)$. 
The theory predicts that resonance energy does not depend on the choice of $\lambda$ and
it is found in practice as well. So $\lambda$ is suitably chosen to optimize the stability 
of the resonant state. It preserves the spectrum of the original potential, only adding a 
discrete BIC at a selected energy.\\

\noindent 
From the constructed family of strictly isospectral potentials $\hat{v}(r;\lambda)$, 
we extract normalizable wave functions $\hat{\psi}_{E}(r;\lambda)$ following the procedures of 
BIC and could effectively calculate the trapping probability of the system within the enlarged 
well-barrier combination as 
\begin{equation}
C(E)={\displaystyle\int}_{r_{a}}^{r_{b}} {[\hat{\psi}_{E}(r^{\prime})]}^{2} dr^{\prime}  \;,
\label{ce}
\end{equation}
where $r_{a}$ and $r_{b}$ are radial distances at the classical turning points $a$, $b$ within the
potential well.
Resonance state which was not apparent in our microscopically constructed potential $v(r)$ now gets prominence in our enlarged well-barrier combinations displayed by the probability plots of $C(E)$. Judicious choice of parameter $\lambda$ only helps in locating the high-lying resonant state while the result remains independent of the choice of $\lambda$ as predicted by the theory. 

\begin{figure}
\includegraphics[scale=0.50]{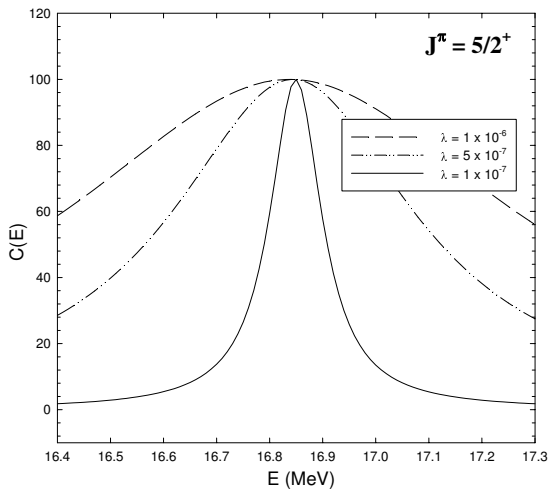}
\caption{\label{fig1}  Probability $C(E)$ as a function of energy $E$ for $\lambda=1 \times 
10^{-6},~5 \times 10^{-7} {\rm ~and~} 1 \times 10^{-7}$ for the $\frac{5}{2}^{+}$ state of  $^9$B.}
\end{figure}

\section{Results}
\noindent 
In finding a solution to the cosmological $^7$Li problem as mentioned above, it was suggested
that the 16.8 MeV state in $^9$B needs to be studied in greater detail~\cite{RI19,KI11,SC11}. 
It is a possibility that substantial amount of $^7$Be 
destruction through resonant reaction with deuteron may lead to this state in $^9$B. Since in
standard Big Bang Nucleosynthesis theory, most $^7$Li is produced in the form of $^7$Be, its
destruction leads to a reduction in $^7$Li abundance. Therefore, inclusion of the resonant 
contribution of the 16.8 MeV state in $^9$B might lead to a solution to the lithium problem.\\

\noindent 
The probability $C(E)$ of the system for $\frac{5}{2}^{+}$ state of $^{9}$B for $\lambda=
1 \times 10^{-6},~5 \times 10^{-7} {\rm ~and~} 1 \times 10^{-7}$ is shown in Fig.~1 where each
curve is normalized to a peak value of 100. The figure clearly indicates the presence of a
resonant state at energy E = 16.84 MeV. This is in excellent agreement to the experimental 
finding of 16.800(10) MeV with a width of 81(5) keV~\cite{SC11}. We calculated the width 
($\Gamma$) of the resonance as 69 keV using WKB method as described in~\cite{DU18}.\\ 

\noindent 
The plot of isospectral potentials $\hat{v}(r;\lambda)$ are shown in Fig. 2 at resonance energy 
$E_{R}$ for the same $\lambda$ values as in Fig. 1 for the $\frac{5}{2}^{+}$ state of $^{9}$B 
along with the original DDM3Y potential and the centrifugal barrier.  The figure shows the
deep well-barrier combination that actually succeeds in trapping the high-lying resonant state. 
The narrow and deep potential wells result in numerical difficulties in computation that indirectly 
limits the choice of $\lambda$ also. The deep and narrow isospectral potentials of depth around
200-300 MeV for the present high-lying resonance at 16.84 MeV can be compared to the isospectral 
potentials of depth around 20-50 MeV for the low-lying resonance of 1.8 MeV in 
$^{15}$Be~\cite{DU18}.\\

\noindent 
The wave functions $\hat{\psi}_{E}(r;\lambda)$ at the resonance energy 16.84 MeV for the same 
$\lambda$ values as in Fig. 1 are shown in Fig. 3. The wave function plots are confirmation of 
the presence of the high-lying resonant state as it has an appreciable amplitude within the 
well. In the asymptotic region, the sinusoidal nature represents a free system after it 
leaks out of the well-barrier combination. The inset of Fig. 3 shows the wave
function plot for $\lambda$ = $5 \times 10^{−7}$ in an expanded scale up to 50 fm. We carried out
similar calculations for a $\frac{3}{2}^{+}$ resonance as shown by the dotted line in Fig. 3.
The low amplitude of such a wave function within the well as well as similar oscillatory
behaviour in the asymptotic region rules out $\frac{3}{2}^{+}$ resonance at this energy.
The observed angular distribution~\cite{SC11} of the 16.8 MeV state is also consistent with the 
$\frac{5}{2}^{+}$ assignment.

\section{Conclusion}

\noindent 
We have generated the wave function of a high-lying resonant state in the
$^9$B nucleus in the SQM framework with a DDM3Y microscopic potential. The resulting
resonance energy and width agree very well with the experimental data. The single parameter
$\lambda$ in the present formalism, appears to enhance resonance effect but it has no role 
in locating the exact resonance energy. Wave functions for different values of this parameter 
are in a sense equivalent as they reproduce the same resonance energy and width of the state.\\

\noindent 
In a nutshell, the SQM is the only procedure by which resonant state wave functions are 
extracted and utilized to effectively reproduce an experimental observable like resonance
width. The present work successfully showed the effectiveness of SQM and its range of 
applicability in the detection of high-lying resonant state and computation of the wave functions. 
We could establish that the present theoretical procedure works extremely well in the 
study of high-lying resonant states in unstable nuclei. Future works may be directed towards
the decay properties of such resonances which are very relevant in the context of nuclear
astrophysics.

\begin{figure}
\includegraphics[scale=0.50]{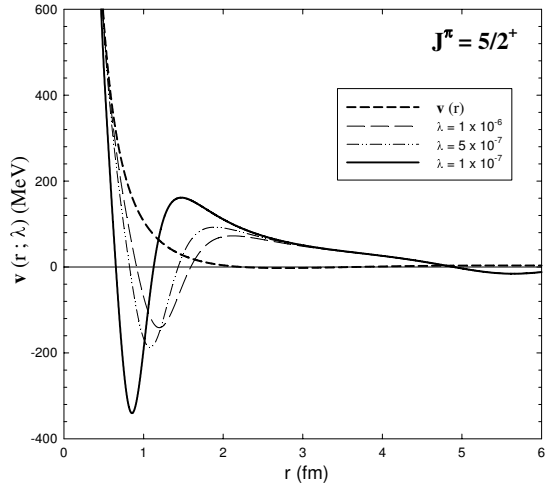}
\caption{\label{fig2} One parameter family of isospectral potentials $V(r; \lambda)$ for 
$\lambda=1 \times 10^{-6},
~5 \times 10^{-7} {\rm ~and~} 1 \times 10^{-7}$ for the $\frac{5}{2}^{+}$ state of $^9$B.}
\end{figure}

\vskip 6. true cm

\newpage
\begin{figure}
\includegraphics[scale=0.50]{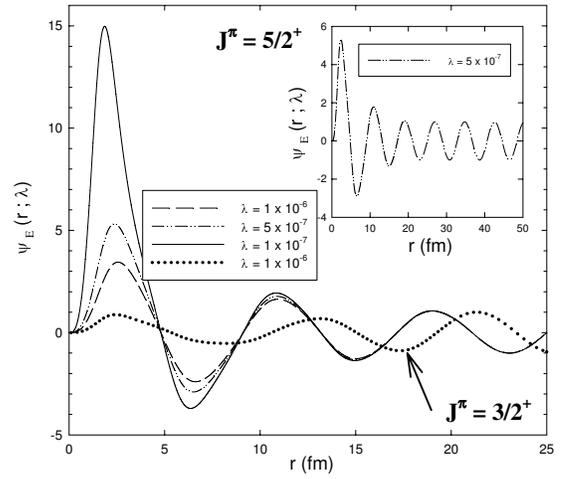}
\caption{\label{fig3} Wave function (in arbitrary units) at the excitation energy of
16.84 MeV for $\lambda = 1 \times 10^{-6},
5 \times 10^{-7} {\rm ~and~} 1 \times 10^{-7}$ for the $\frac{5}{2}^{+}$ state 
of  $^9$B. The inset shows the wave function plot for $\lambda = 5 \times 10^{-7}$ in an 
expanded scale up to 50 fm.}
\end{figure}

\end{document}